%% file: main.tex
\documentclass[9pt,conference]{IEEEtran}

\IEEEoverridecommandlockouts

\usepackage{amsmath}
\usepackage{comment}
\usepackage{fancyhdr}
\usepackage[normalem]{ulem}
\usepackage{color}
\usepackage{graphicx}
\usepackage{courier}
\usepackage{array}

\usepackage{amssymb}
\usepackage{subfig}
\usepackage[english]{babel}
\usepackage{soul,color,pifont,cite,tikz,balance}
\usepackage{url}
\usepackage{multirow}

\begin{document}
\title{\huge Cost-Aware Exploration for Chiplet-Based Architecture \\with Advanced Packaging Technologies 
\thanks{This paper is presented in HiPChips Chiplet Workshop, co-located with International Symposium on Computer Architecture (ISCA), 2022}

\vspace{-10pt}

}

\author{Tianqi~Tang (tianqi\_tang@ucsb.edu), Yuan~Xie\\University of California, Santa Barbara \vspace{-10pt}}
\IEEEtitleabstractindextext{%
\begin{abstract}
The chiplet-based System-in-Package~(SiP) technology enables more design flexibility via various inter-chiplet connection and heterogeneous integration.
However, it is not known how to convert such flexibility into cost efficiency, which is critical when making a design decision.
In this paper, we develop an analytical cost model that can estimate the cost of the 2.5D chiplet-based SiP systems under various interconnection options and technology nodes.
We conducted two case studies using our cost model to explore the cost characteristics of the 2.5D chiplet-based SiP system.
Based on the case studies, we made several observations on the interposer selection, design partition granularity, and technology node adoption for cost-efficient chiplet-based SiP design.

\end{abstract}

 }

\maketitle

\thispagestyle{plain}
\pagestyle{plain}

\IEEEdisplaynontitleabstractindextext
\IEEEpeerreviewmaketitle


\input{intro.tex}
\input{cost_model.tex}

\input{case_study.tex}
\input{conclusion.tex}

{\scriptsize
\bibliographystyle{IEEEtran}
\bibliography{ref.bib}}
\end{document}

%% file: intro.tex
\section{Introduction}
Recently, the ``chiplet''-based System-in-Package (SiP) becomes the potential replacement of the conventional System-on-Chip (SoC) which suffers from the increasing complexity and cost of new technology nodes.
The SiP breaks a monolithic die based 2D SoC into multiple smaller pieces and keep them in the same package.
These dies can be of different functionalities, hybrid technology nodes, and/or from multiple IP designers.
Such a heterogeneous integration greatly reduces the design complexity of each die due to IP reuse.
So far, the majority of the chiplet system research focuses on either demonstrating the chiplet prototypes~\cite{zimmer20200,naffziger20202,naffziger2021pioneering,arunkumar2017mcm}; or studying the workload-aware dataflow~\cite{tan2021nn}, and the network-on-package~\cite{goyal2020neksus,coskun2020cross,yin2018modular} under the performance, energy, and thermal constraints. 
However, little attention has been given to the cost, which will become a critical factor whether to adopt the chiplet/SiP in the design.
All the potential advantages of the chiplet related technologies ultimately have to be translated into cost savings when evaluating a design strategy.

The analytical cost model can be the solution to estimate the cost and guide the early-stage cost-aware design. 
Previously, the analytical cost model has been used in evaluating the TSV-based 3D architecture~\cite{zhao2010cost} or the silicon interposer based 2.5D integrated system~\cite{stow2016cost,stow2017cost}. 
These works cannot be directly adopted because several new design choices emerge in the chiplet-based SiP design and are not yet well explored, including the inter-chiplet connection, the homogeneous/heterogeneous integration, make up a large and complex design space that is not yet well explored. 

In this paper, we make, to our best, the first attempt to build a cost model for chiplet-based SiP design space exploration.
With the input of the system scale (e.g., the transistor count of compute die, memory cell count, and other statistics), the partition granularity (e.g., the die count), and the technology node, our cost model will first 
translate the system scale into the area and the number of wiring layers. 
The cost breakdown of the die, the bonding, and the package is then calculated based on these data. 
With the collected data of transistor density as well as the wiring pitches, our cost model is able to support the 2.5D Silicon interposer, the 2.5D organic interposer, and MCM under the technology nodes ranging from 28nm to 5nm to 
support the heterogeneous integration. Using the proposed cost model, we perform two case studies to explore the cost characteristics of the 2.5D chiplet-based SiP system; and made several observations on the interposer selection, design partition granularity, and technology node adoption for cost-efficient chiplet-based SiP design.

%% file: cost_model.tex
\section{Analytical Cost Model for Chiplet System with Interposer and MCM-based Integration}
\label{sec:cost_model}

Our analytical cost model is made up of the manufacture cost of each die/interposer, the bonding cost, and the package cost; it is able to model the chiplet system with the silicon interposer, organic interposer, and MCM-based integration. The manufacture cost of single die and silicon interposer has been well studied in previous work \cite{stow2017cost}. For space limitation, this section focuses on the  manufacture cost of organic interposer, the bonding and package cost.

\subsection{Manufacture Cost Model of An Individual Organic Interposer}
\label{subsec:die_cost}

 The organic interposer is built in a panel form (i.e., a large square or rectangle), different from die and silicon interposer, which are built in the wafer form,rather than a wafer form (i.e., a round plate). Eq. \ref{eq:org_int_cost}. shows the cost ($C_{org\_int}$) and yield ($Y_{org\_int}$) of an organic interposer.
\begin{equation}
\label{eq:org_int_cost}
\begin{split}
    C_{org\_int} &= \frac{C_{panel}}{N_{org\_int}} = \frac{C_{panel}}{A_{panel}/ A_{org\_int}} \\
    Y_{org\_int} &= Y_{panel}\times(1+\frac{A_{org\_int}D_0}{\alpha})^{-\alpha}
\end{split}
\end{equation}
Here $C_{panel}$ and $N_{org\_int}$ are the panel cost and the number of organic interposers obtained from a panel.
$A_{panel}$ and $A_{org\_int}$ are the area of the panel and the organic interposer,  respectively. 
$Y_{panel}$, $D_0$, $\alpha$ are the yield, defect density, and defect clustering parameter of the interposer panel, respectively. They are collected from ICKnowledge \cite{icknowledge} and are determined by the technology node. 

\subsection{Bonding Cost Model}
\label{subsec:manu_cost}

The unpackaged interposer-based chiplet system is made up of $n$ functional dies on top and one interposer die at the bottom; while the MCM based chiplet system directly deploys $n$ functional dies onto the organic substrate. The bonding cost of the unpackaged chiplet system is calculated by Eq. \ref{eq:wafer_cost_interposer} and Eq. \ref{eq:wafer_cost_mcm}, which represent the interposer-based chiplet system ($C_{int\_2.5D}$) and the MCM-based chiplet system ($C_{MCM\_2.5D}$), respectively.
\begin{equation}
\label{eq:wafer_cost_interposer}
C_{int\_2.5D} = \frac{\frac{C_{int}}{Y_{int}}+\sum_{i=1}^{n}{(\frac{C_{die}(i)}{Y_{die}(i)}+C_{bond}(i))}}{\prod_{i=2}^{n}{Y_{bond}(i)}}
\end{equation}
\begin{equation}
\label{eq:wafer_cost_mcm}
C_{MCM\_2.5D} = \frac{\sum_{i=1}^{n}{(\frac{C_{die}(i)}{Y_{die}(i)}+C_{bond}(i))}}{\prod_{i=2}^{n}{Y_{bond}(i)}}
\end{equation}
Here $n$ is the number of function dies.
$C_{die}(i)$, and $Y_{die}(i)$ are the manufacturing cost and yield for the $i^{th}$ functional die.
$C_{bond}(i)$ and $Y_{bond}(i)$ are the bonding cost and yield for the $i^{th}$ die.
In most cases each functional die is assumed to have the same bonding cost and yield.
$C_{int}$ and $Y_{int}$ are the manufacturing cost and the yield for the interposer. Comparing Eq. \ref{eq:wafer_cost_interposer} and Eq. \ref{eq:wafer_cost_mcm}, the cost of the interposer based chiplet system has an extra term of $\frac{C_{int}}{Y_{int}}$, which depends on the material of the interposer.




\subsection{Package Cost Model}
\label{subsec:package_cost}
The package cost depends on the type of package. Flip chip based organic substrate can be used in both interposer based and MCM based chiplet system. For simplification 
we use the flip chip based organic substrate to showcase the package cost model. Similar cost models can also be used in other package types. 
In addition to the package type, the package cost is also determined by three factors, i.e., the package area, the layer number of package (\#core and \#buildup), and the pin count.

Given the numbers of core layers and the build-up layers, we collect the package cost under different combinations of substrate area and pin count from ICKnowledge~\cite{icknowledge}. With the collected data, we derive an empirical function of the package cost ($C_P$) with respect to the combination of the substrate area ($A_{sub}$) and the number of pins ($N_{pin}$) in Eq.\ref{eq:package_cost}. $\mu_{0}, \mu_{1}, \mu_{2}$ are the regression parameters determined by the numbers of core layers and the build-up layers.
\begin{equation}
\label{eq:package_cost}
    C_P =  \mu_0  A_{sub}  + \mu_1 N_{pin} + \mu_2
\end{equation}

\begin{figure}[t]
    \centering
    \includegraphics[width=8.6cm]{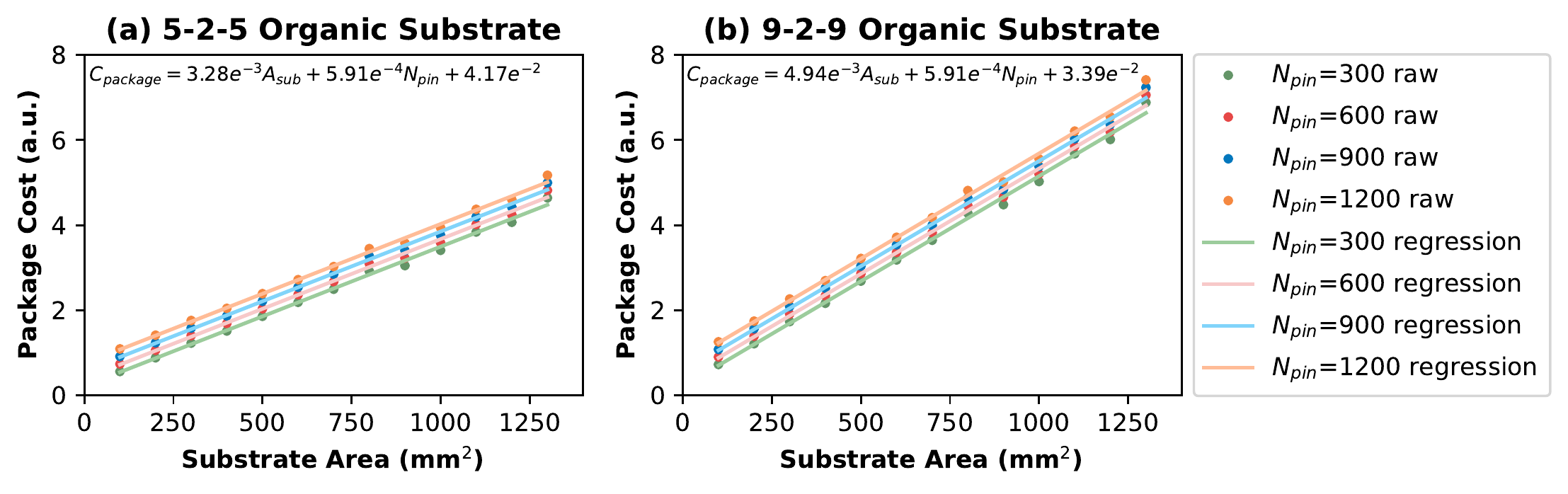}\vspace{-5pt}

    \caption{Package cost of organic substrates}
    \vspace{-10pt}
    \label{fig:package_cost_mcm}
\end{figure}

Fig. \ref{fig:package_cost_mcm} shows the regressive package cost model of the organic substrates of two different configurations (a) 2 core layers and 5 build-up layers; and (b) 2 core layers and 9 build-up layers. The result shows a satisfying linearity of the developed regressive model. We will use this regressive package model in the follow-up case studies.

%% file: case_study.tex
\section{Case Study}
\label{sec:case_study}

In this section, we leverage the developed cost model 
to conduct two case studies and showcase the cost characteristics of the chiplet based architecture, i.e. 
a  heterogeneous  multi-chiplet  system with HBM stacks (Sec. \ref{subsec:cost_model_heter_hbm}), and
a heterogeneous multi-chiplet system which partitions the core components and IO components onto different chiplets under different processes (Sec. \ref{subsec:cost_model_heter_io}). 

\begin{table}[t]
\scriptsize
\caption{Switching points of organic interposer/MCM based chiplet systems under different technology nodes}
\begin{tabular}{|c|c|c|c|c|c|c|c|}
\hline
\begin{tabular}[c]{@{}c@{}}Tech \\ Node\end{tabular}                          & \begin{tabular}[c]{@{}c@{}}Chiplet\\ Type\end{tabular} & 7nm   & 10nm  & 12nm  & 16nm & 20nm & 28nm \\ \hline
\multirow{2}{*}{\begin{tabular}[c]{@{}c@{}}2D Area\\ (mm$^2$)\end{tabular}}      & Org 2.5D                                               & 178   & 191   & 264   & 279  & 313  & 479  \\ \cline{2-8} 
                                                                              & MCM                                                    & 119   & 120   & 126   & 128  & 131  & 149  \\ \hline
\multirow{2}{*}{\begin{tabular}[c]{@{}c@{}}Tx Count\\ (Billion)\end{tabular}} & Org2.5D                                                & 17.17 & 10.71 & 11.22 & 7.66 & 6.12 & 5.62 \\ \cline{2-8} 
                                                                              & MCM                                                    & 11.48 & 6.73  & 5.36  & 3.51 & 2.56 & 1.75 \\ \hline
\end{tabular}
\vspace{-10pt}
\label{tab:switching_point_homo}
\end{table}


\subsection{Heterogeneous Chiplet System with HBM Stacks}
\label{subsec:cost_model_heter_hbm}

The heterogeneous chiplet system with high bandwidth requirement integrates the core dies and the HBM stacks onto the silicon interposer or the organic interposer. Each HBM stack is made up of a base die at the bottom and several layers of memory dies atop \cite{kim2014hbm}. For the base die, the area of a 1024-bit signal interface is mainly determined by the pitch width of the microbumps. We set the pitch width as 45$\mu \textrm{m}$ for the  silicon interposer and 110$\mu \textrm{m}$ for the  organic interposer. We exclude MCM in this case study because the C4 bump pitch is too large to place the whole signal and I/O interface under the area constraint of existing HBMs\footnote{According to public data \cite{kim2014hbm}, HBM1 takes the area of $5.48 \textrm{mm} \times 7.29 \textrm{mm}$. This implies that when the bump pitch width is larger than $197\mu \textrm{m}$, the 1024-bit I/O interface will use up the overall area budget of HBM stack.}. 

\begin{figure*}[t]
    \centering
    \includegraphics[width=18cm]{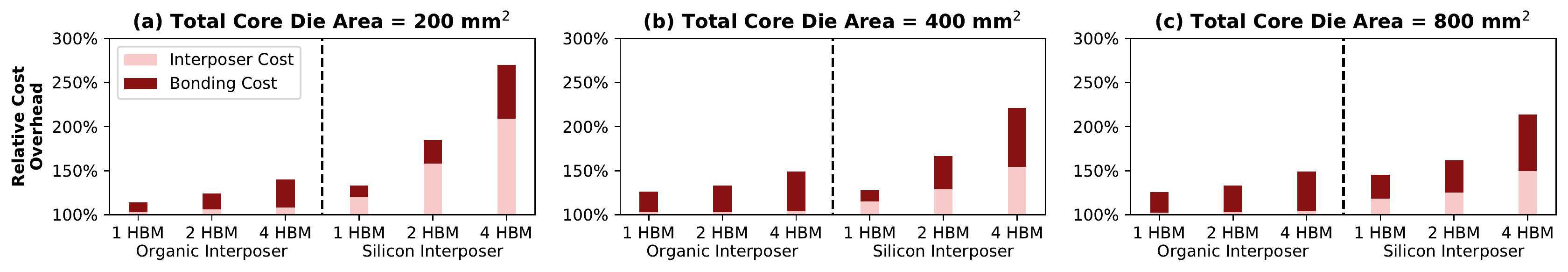}
    \vspace{-8pt}
    \caption{Relative cost overhead of heterogeneous chiplet system with HBM stacks of different scales under 7nm process}\vspace{-10pt}
    \label{fig:heter_hbm}
\end{figure*}

This case study aims to study the extra cost introduced by HBM stacks. Fig. \ref{fig:heter_hbm} visualizes the relative cost breakdown of interposer \footnote{The cost of interposer considers the area of HBM stacks. Yet the cost of memory stacks is not calculated due to the lack of data.} and bonding, where the manufacture cost of core dies as the 100\% base unit. The three subfigures show the system scales of 200mm$^2$, 400mm$^2$, and 800mm$^2$ under 7nm process respectively. We observe that the organic interposer based chiplet system introduces less than 50\% overhead for HBM stacks and the bonding yield takes the majority of the overhead. While for the silicon interposer based chiplet system, the relative cost overhead is much larger.


\begin{figure*}[t]
    \centering
    \includegraphics[width=18cm]{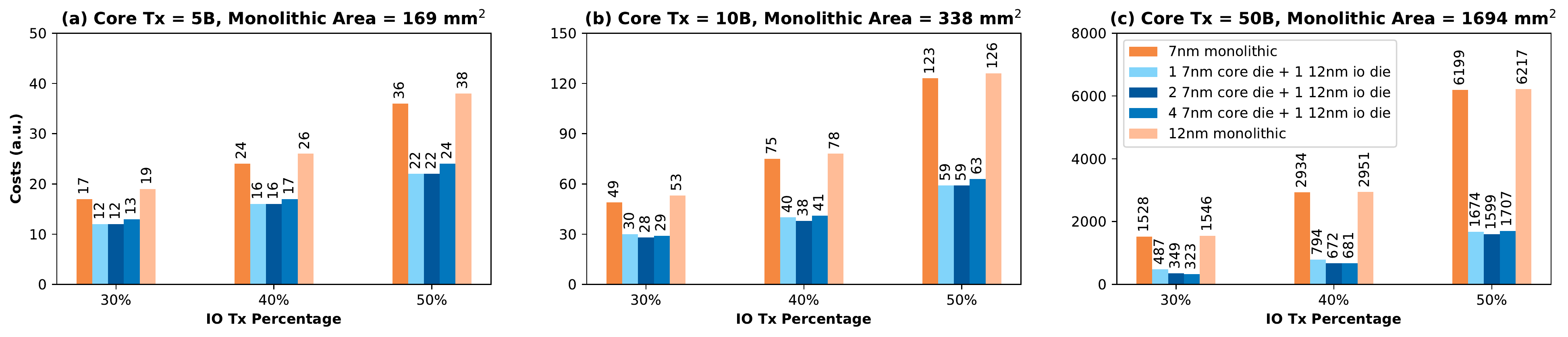}\vspace{-8pt}
    \caption{Cost of MCM based heterogeneous chiplet system with hybrid processes under different system scales}\vspace{-10pt}
    \label{fig:heter_exp_cropped}
\end{figure*}

\subsection{Heterogeneous Chiplet System with Hybrid Processes}
\label{subsec:cost_model_heter_io}

Considering the technology scaling of I/O and other peripheral circuits are much slower than that of the compute logic and the on-chip memory \cite{wong2020density}, the cost efficiency may potentially increase if different components are assigned to different dies and are implemented in different technology nodes. Inspired by AMD's EYPC2 CPU \cite{naffziger20202,naffziger2021pioneering}, this case study explores the heterogeneous architecture where the cores and the IO peripheral circuits are split into different dies and  implemented in different technology nodes. As shown in Fig. \ref{fig:heter_exp_cropped}, we study the MCM based chiplet system under different system scales and different I/O proportion. To quantitatively explore the cost efficiency of the heterogeneous chiplet system, we compare the cost of a 7nm monolithic system, a 12nm monolithic system, and three hybrid systems with different numbers of core dies where the core dies and the I/O die are respectively assumed at 7nm and 12nm technology nodes. The three subfigures of Fig. \ref{fig:heter_exp_cropped} respectively show the circumstances when the numbers of transistors on the core die are 5 billion, 10 billion, and 50 billion. And each subfigure includes the proportions of I/O circuits in the range of \{30\%, 40\%, 50\%\}. 

We find that the hybrid system which partitions the core logics and the I/O circuits into different dies achieves salient benefits on cost efficiency compared with the monolithic systems under both the mature and advanced technology nodes. Moreover, as the scale of the chiplet system gets larger, both the cost efficiency of the hybrid system and the optimal number of core dies increase. For the three different system scales in Fig. \ref{fig:heter_exp_cropped}, the optimal chiplet system respectively achieves 34\%, 48\%, and 77\% cost efficiency improvement compared with the 7nm monolithic counterpart, and the optimal numbers of core dies are respectively 2, 4, and 8.


%% file: conclusion.tex
\section{Conclusion}
\label{sec:conclusion}

In this paper, we build an analytical cost model for the 2.5D chiplet system under various interconnection options and technology nodes. We conduct a series of case studies to explore the cost characteristics under both homogeneous and heterogeneous scenarios. By analysing the case study results, we made several observations on the interposer selection, design partition granularity,  and  hybrid technology  node  adoption  for  the cost-efficient  chiplet-based  SiP  design.

Besides the manufacture cost and the package cost discussed in this paper, the testing cost and the cooling cost also play an important role in the early-stage design decision. Meanwhile, the exploration would be more comprehensive when considering the performance, power, area, and cost simultaneously. All these will be explored in our future work.


%% file: main.bbl
\begin{thebibliography}{10}
\providecommand{\url}[1]{#1}
\csname url@samestyle\endcsname
\providecommand{\newblock}{\relax}
\providecommand{\bibinfo}[2]{#2}
\providecommand{\BIBentrySTDinterwordspacing}{\spaceskip=0pt\relax}
\providecommand{\BIBentryALTinterwordstretchfactor}{4}
\providecommand{\BIBentryALTinterwordspacing}{\spaceskip=\fontdimen2\font plus
\BIBentryALTinterwordstretchfactor\fontdimen3\font minus
  \fontdimen4\font\relax}
\providecommand{\BIBforeignlanguage}[2]{{%
\expandafter\ifx\csname l@#1\endcsname\relax
\typeout{** WARNING: IEEEtran.bst: No hyphenation pattern has been}%
\typeout{** loaded for the language `#1'. Using the pattern for}%
\typeout{** the default language instead.}%
\else
\language=\csname l@#1\endcsname
\fi
#2}}
\providecommand{\BIBdecl}{\relax}
\BIBdecl

\bibitem{zimmer20200}
B.~Zimmer \emph{et~al.}, ``A 0.32--128 tops, scalable multi-chip-module-based
  deep neural network inference accelerator with ground-referenced signaling in
  16 nm,'' \emph{IEEE Journal of Solid-State Circuits}, vol.~55, no.~4, pp.
  920--932, 2020.

\bibitem{naffziger20202}
S.~Naffziger \emph{et~al.}, ``Amd chiplet architecture for high-performance
  server and desktop products,'' in \emph{2020 IEEE International Solid-State
  Circuits Conference-(ISSCC)}.\hskip 1em plus 0.5em minus 0.4em\relax IEEE,
  2020, pp. 44--45.

\bibitem{naffziger2021pioneering}
S.~Naffziger, N.~Beck, T.~Burd, K.~Lepak, G.~H. Loh, M.~Subramony, and
  S.~White, ``Pioneering chiplet technology and design for the amd epyc™ and
  ryzen™ processor families: Industrial product,'' in \emph{2021 ACM/IEEE
  48th Annual International Symposium on Computer Architecture (ISCA)}.\hskip
  1em plus 0.5em minus 0.4em\relax IEEE, 2021, pp. 57--70.

\bibitem{arunkumar2017mcm}
A.~Arunkumar \emph{et~al.}, ``Mcm-gpu: Multi-chip-module gpus for continued
  performance scalability,'' \emph{ACM SIGARCH Computer Architecture News},
  vol.~45, no.~2, pp. 320--332, 2017.

\bibitem{tan2021nn}
Z.~Tan \emph{et~al.}, ``Nn-baton: Dnn workload orchestration and chiplet
  granularity exploration for multichip accelerators,'' in \emph{2021 ACM/IEEE
  48th Annual International Symposium on Computer Architecture (ISCA)}.\hskip
  1em plus 0.5em minus 0.4em\relax IEEE, 2021, pp. 1013--1026.

\bibitem{goyal2020neksus}
V.~Goyal \emph{et~al.}, ``Neksus: An interconnect for heterogeneous
  system-in-package architectures,'' in \emph{2020 IEEE International Parallel
  and Distributed Processing Symposium (IPDPS)}.\hskip 1em plus 0.5em minus
  0.4em\relax IEEE, 2020, pp. 12--21.

\bibitem{coskun2020cross}
A.~Coskun \emph{et~al.}, ``Cross-layer co-optimization of network design and
  chiplet placement in 2.5-d systems,'' \emph{IEEE Transactions on
  Computer-Aided Design of Integrated Circuits and Systems}, vol.~39, no.~12,
  pp. 5183--5196, 2020.

\bibitem{yin2018modular}
J.~Yin \emph{et~al.}, ``Modular routing design for chiplet-based systems,'' in
  \emph{2018 ACM/IEEE 45th Annual International Symposium on Computer
  Architecture (ISCA)}.\hskip 1em plus 0.5em minus 0.4em\relax IEEE, 2018, pp.
  726--738.

\bibitem{zhao2010cost}
J.~Zhao \emph{et~al.}, ``Cost-aware three-dimensional (3d) many-core
  multiprocessor design,'' in \emph{Design Automation Conference}.\hskip 1em
  plus 0.5em minus 0.4em\relax IEEE, 2010, pp. 126--131.

\bibitem{stow2016cost}
D.~Stow \emph{et~al.}, ``Cost analysis and cost-driven ip reuse methodology for
  soc design based on 2.5 d/3d integration,'' in \emph{Proceedings of the 35th
  International Conference on Computer-Aided Design}, 2016, pp. 1--6.

\bibitem{stow2017cost}
------, ``Cost-effective design of scalable high-performance systems using
  active and passive interposers,'' in \emph{2017 IEEE/ACM International
  Conference on Computer-Aided Design (ICCAD)}.\hskip 1em plus 0.5em minus
  0.4em\relax IEEE, 2017, pp. 728--735.

\bibitem{icknowledge}
``Ic cost and price model, icknowledge llc,''
  \url{https://www.icknowledge.com/products/icmodel.html}, accessed: Sept 8,
  2021.

\bibitem{kim2014hbm}
J.~Kim and Y.~Kim, ``Hbm: Memory solution for bandwidth-hungry processors,'' in
  \emph{2014 IEEE Hot Chips 26 Symposium (HCS)}.\hskip 1em plus 0.5em minus
  0.4em\relax IEEE, 2014, pp. 1--24.

\bibitem{wong2020density}
H.-S.~P. Wong \emph{et~al.}, ``A density metric for semiconductor technology
  [point of view],'' \emph{Proceedings of the IEEE}, vol. 108, no.~4, pp.
  478--482, 2020.

\end{thebibliography}
